# Surrogate Modeling for Neutron Transport: A Neural Operator Approach

Md Hossain Sahadath, Qiyun Cheng, Shaowu Pan, Wei Ji[1]

Department of Mechanical, Aerospace, and Nuclear Engineering, Rensselaer Polytechnic Institute, 110 8th Street, Troy, NY

## ABSTRACT

This work introduces a neural operator based surrogate modeling framework for neutron transport computation. Two architectures, the Deep Operator Network (DeepONet) and the Fourier Neural Operator (FNO), were trained for fixed source problems to learn the mapping from anisotropic neutron sources, $Q(x, \mu)$, to the corresponding angular fluxes, $\psi(x, \mu)$, in a one-dimensional slab geometry. Three distinct models were trained for each neural operator, corresponding to different scattering ratios ($c = 0.1, 0.5, \& 1.0$), providing insight into their performance across distinct transport regimes (absorption-dominated, moderate, and scattering-dominated). The models were subsequently evaluated on a wide range of previously unseen source configurations, demonstrating that FNO generally achieves higher predictive accuracy, while DeepONet offers greater computational efficiency. Both models offered significant speedups that become increasingly pronounced as the scattering ratio increases, requiring <0.3% of the runtime of a conventional $S_N$ solver. The surrogate models were further incorporated into the $S_N$ k-eigenvalue solver, replacing the computationally intensive transport sweep loop with a single forward pass. Across varying fission cross sections and spatial-angular grids, both neural operator solvers reproduced reference eigenvalues with deviations up to 135 pcm for DeepONet and 112 pcm for FNO, while reducing runtime to <0.1% of that of the $S_N$ solver on relatively fine grids. These results demonstrate the strong potential of neural operator frameworks as accurate, efficient, and generalizable surrogates for neutron transport, paving the way for real-time digital twin applications and repeated evaluations, such as in design optimization.

---

[1] Corresponding Author, Email: jiw2@rpi.edu

## I. INTRODUCTION

A precise description of neutron behavior in space, angle, and energy is crucial for reactor design, criticality safety, shielding evaluations, and multiphysics analyses. These behaviors are governed by the neutron transport equation (NTE), whose numerical solution reveals detailed neutron interaction patterns within the systems [1, 2]. Although deterministic and Monte Carlo solvers can produce highly accurate results, they often incur considerable computational cost, which becomes a major constraint for applications requiring rapid or repeated model evaluations, such as optimization studies or uncertainty quantification. The necessity for computational efficiency is even more pronounced in digital-twin frameworks, where real-time or near-real-time predictions are needed to support operational decision making [3]. Thus, obtaining high-fidelity transport solutions at reduced cost remains an ongoing challenge. Recent progress in deep learning, especially the development of physics-informed neural networks (PINNs) and their extensions, has spurred interest in applying these methods to the NTE and other neutron transport problems [4–6]. While promising, these methods typically demonstrate limited generalization to unseen scenarios, particularly when the underlying physical configuration is modified. Both classical numerical solvers and machine learning models are generally formulated for fixed physical configurations [7], meaning that any change in model inputs usually requires a new simulation or complete retraining, resulting in significant computational overhead and constrained flexibility.

Neural operator methodologies have emerged as a compelling alternative for constructing efficient surrogate models for partial differential equations (PDEs) [8]. Architectures such as Deep Operator Network (DeepONet) [9] and Fourier Neural Operator (FNO) [10] are capable of learning solution operators that map entire input functions to output functions in infinite dimensional spaces. Unlike traditional neural networks that learn discrete pointwise relations, neural operators learn continuous function-to-function mappings, enabling them to more effectively capture the underlying physics and generalize more reliably than other machine learning approaches across unseen scenarios without retraining. These models have demonstrated strong performance across a range of scientific and engineering applications, including material fracture prediction [11], thermal transport modeling [12], multiphysics problems [13], multiphase flow simulations [14], and fusion plasma dynamics [15]. Despite this progress, applications of neural operators to neutron transport remain limited [16-19].

Building on our earlier efforts [16-18], in which we developed neural operator based surrogate model to map isotropic neutron source $S(x)$ to the scalar flux $\phi(x)$, the present work extends the methodology to neutron transport problem involving anisotropic sources. Neural operator based surrogate models are constructed to learn the mapping from an anisotropic source term $Q(x,\mu)$ to the corresponding angular flux $\psi(x,\mu)$. Using training data generated from high-fidelity fixed source NTE simulations with fixed macroscopic cross sections, both DeepONet and FNO architectures are trained to approximate the underlying transport operator. Once trained, the models are employed to solve the fixed source NTE for a broad range of anisotropic source configurations that are unseen during training. In addition to the fixed source problem, the DeepONet and FNO models were subsequently integrated into the eigenvalue formulations of the NTE. In both cases, fixed source and eigenvalue NTE, they offer super-fast evaluations, as no transport sweep is required. Both models demonstrate the capability to provide rapid and accurate evaluations of the angular flux, thereby enabling an efficient, generalizable, and reliable solution framework for the NTE. Moreover, accurately capturing the inherently multiphysics behavior of nuclear reactors requires fast neutron transport solvers to enable efficient coupling of neutronics and thermal hydraulics analyses [20-22]. In this context, neural operators may offer a promising surrogate modeling approach for such coupled simulations.

## II. METHODOLOGY

### II.A. NTE for Fixed Source Problem

This work investigates the fixed source neutron transport problem in a one dimensional slab geometry, illustrated in Figure 1. The slab spans 10 cm and is enclosed by vacuum boundaries that impose zero incoming current. The absorption, scattering, and total cross sections are taken to be spatially uniform, and a prescribed internal source is distributed throughout the slab.

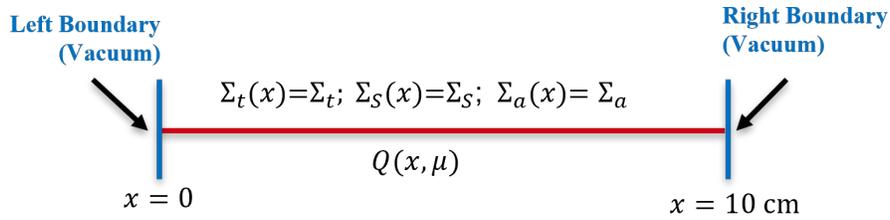

Figure 1. 1D Slab considered for the fixed source NTE.

The steady-state one-group NTE for this one-dimensional slab geometry is expressed as follows:

$$\mu \frac{\partial \psi(x,\mu)}{\partial x} + \Sigma_t \, \psi(x,\mu) = \int_{-1}^{1} \frac{1}{2} [\Sigma_{S0} + 3\mu\mu' \Sigma_{S1}] \, \psi(x,\mu') d\mu' + Q(x,\mu);$$

$$0 < x < X, \ -1 \leq \mu \leq 1 \tag{1}$$

$$\psi(0,\mu) = 0, \ 0 \leq \mu \leq 1$$

$$\psi(X,\mu) = 0, \ -1 \leq \mu \leq 0$$

Here, $\mu$ denotes the cosine of the scattering angle, $\psi(x,\mu)$ represents the angular neutron flux, $\Sigma_t$ is the total cross section, $\Sigma_{S0}$ and $\Sigma_{S1}$ are the zeroth and first Legendre moments of the scattering cross section, and $Q(x,\mu)$ is the internal neutron source. The left-hand side represents neutron losses, while the right-hand side captures the effects of in-scattering and the prescribed internal source. With the cross sections ($\Sigma_t, \Sigma_{S0}$ and $\Sigma_{S1}$), held constant, variations in the internal source $Q(x,\mu)$ lead to different angular flux outcomes $\psi(x,\mu)$. The first goal of this study is to develop a neural operator surrogate that can efficiently learn the mapping from $Q(x,\mu)$ to $\psi(x,\mu)$. Accordingly, DeepONet and FNO networks were trained using datasets encompassing diverse source distributions and their computed angular flux solutions.

## II.B. Deep Operator Network (DeepONet)

DeepONet is grounded the Universal Approximation Theorem for Operators [8, 23], which states that neural networks can approximate nonlinear operators acting on functions. In this framework, the operator $G: Q(x,\mu) \mapsto \psi(x,\mu)$ represented through two subnetworks: the BranchNet and the TrunkNet. The BranchNet encodes the discretized $Q(x,\mu)$ into a latent vector $\boldsymbol{b_N} = [b_1, b_2, \ldots, b_p]$, while the TrunkNet maps query coordinates $(x,\mu)$ into a mode vector $\boldsymbol{t_N} = [t_1(x,\mu), t_2(x,\mu), \ldots, t_p(x,\mu)]$ as shown in Figure 2. The predicted $\psi(x,\mu)$ is then computed as the inner product of these two vectors:

$$\psi(x,\mu) \approx \boldsymbol{b_N} \cdot \boldsymbol{t_N^T} = \sum_{i=1}^{p} b_i t_i(x,\mu) \qquad (2)$$

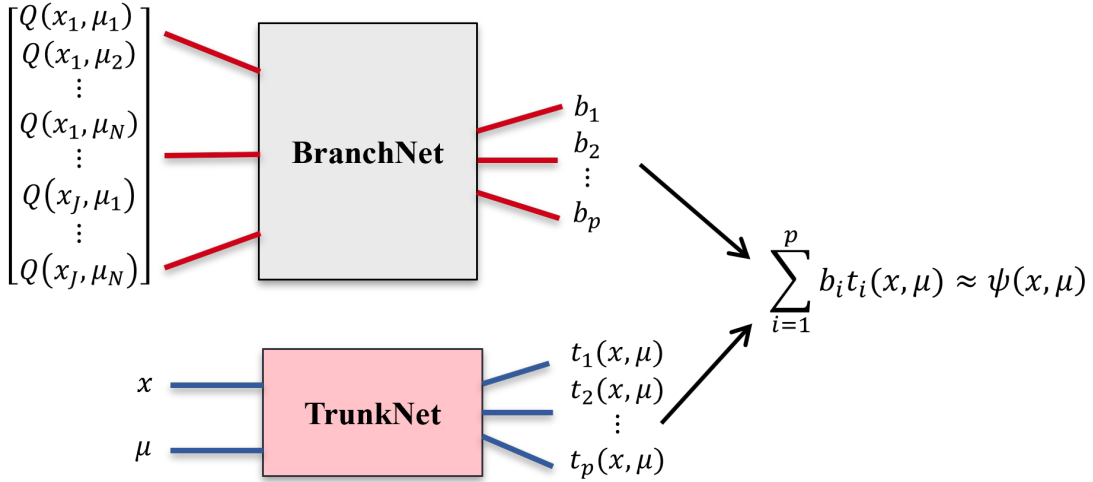

Figure 2. Architecture of DeepONet.

By leveraging this structure, DeepONet can effectively capture the operator, providing rapid and precise flux estimates for new source profiles over arbitrary spatial and angular coordinates, facilitating real-time transport calculations.

**II.C. Fourier Neural Operator (FNO)**

The FNO provides a data-driven approach for approximating solution operators of partial differential equations [10]. In this study, it is employed to learn the mapping from the internal source $Q(x,\mu)$ to the angular flux $\psi(x,\mu)$. The discretized source is first lifted into a higher-dimensional latent space using a shallow fully connected network, producing $\tilde{Q}$, which is then propagated through a sequence of Fourier layers shown in in Figure 3. Each Fourier layer first applies an FFT to transfer the input into Fourier space, then uses a learnable linear layer to map the Fourier coefficients (to solution space), after which high-frequency modes may be truncated. The data is then transformed back to the spatial domain via an inverse FFT. The FFT/IFFT are fixed computations; the linear mapping is the only learnable component of the layer. In parallel, a pointwise linear transformation is applied directly in the spatial domain, and the results from two pathways are merged through a nonlinear activation. After $F$ such layers, the output representation $\tilde{Q}_F$ is projected to the target dimension to obtain $\psi(x,\mu)$. A key advantage of the FNO is its ability to perform zero-shot super-resolution, enabling models trained on coarse data to be evaluated on finer grids without retraining [10].

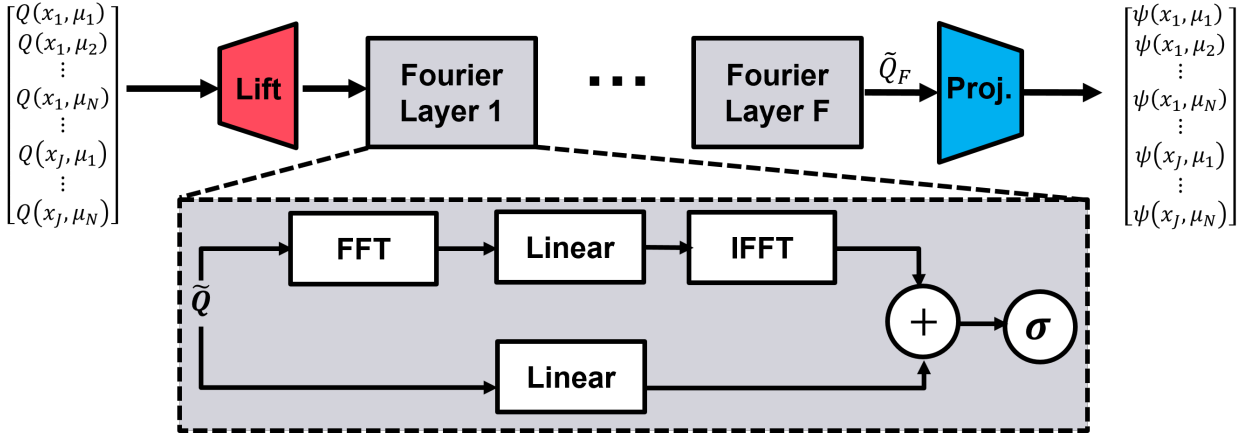

Figure 3. Architecture of FNO.

**II.D. Model Training**

Datasets for training the DeepONet and FNO models were generated by solving Eq. (1) with spatial and angular discretization. The spatial region was divided into $J$ equally spaced cells of width $h_j = 0.1$ cm, and a cell-averaged finite difference scheme was used for spatial discretization [1]. The $j^{th}$ cell spans from $x_{j-\frac{1}{2}}$ to $x_{j+\frac{1}{2}}$, with its width and center defined in Eq. (3).

$$h_j = x_{j+\frac{1}{2}} - x_{j-\frac{1}{2}} \text{ and } x_j = \frac{1}{2}\left(x_{j+\frac{1}{2}} + x_{j-\frac{1}{2}}\right) \tag{3}$$

The quantity $p_j$ represents the cell-averaged value of $p(x)$ across the $j^{th}$ cell and is given by

$$p_j = \frac{1}{h_j} \int_{x_{j-\frac{1}{2}}}^{x_{j+\frac{1}{2}}} p(x)dx \qquad (4)$$

Angular discretization was performed using the discrete ordinates ($S_N$) scheme [1] with Gauss–Legendre quadrature and $N = 16$ angles. Incorporating these spatial and angular discretization, Eq. (1) takes the following form:

$$\frac{\mu_n}{h_j}\left(\psi_{n,j+\frac{1}{2}} - \psi_{n,j-\frac{1}{2}}\right) + \Sigma_{tj}\psi_{n,j} = \frac{1}{2}\left(\Sigma_{S0,j}\varphi_{o,j} + 3\mu_n\Sigma_{S1,j}\varphi_{1,j} + Q_{n,j}\right)$$

$$1 \leq n \leq N,\ 1 \leq j \leq J$$

$$\psi_{n,\frac{1}{2}} = 0,\ \frac{N}{2}+1 \leq n \leq N, \mu_n > 0$$

$$\psi_{n,J+\frac{1}{2}} = 0,\ 1 \leq n \leq \frac{N}{2}, \mu_n < 0$$

(5)

In this formulation, $n$ indexes the angular bin and $j$ the spatial cell. The value $\mu_n$ corresponds to the cosine of the scattering angle for direction $n$. The angular fluxes at the right and left faces of cell $j$ are $\psi_{n,j+\frac{1}{2}}$ and $\psi_{n,j-\frac{1}{2}}$, respectively. The total cross section is $\Sigma_{tj}$, and the cell-averaged flux $\psi_{n,j}$ is specified in Eq. (4). The zeroth and first Legendre moments of the scattering cross section are $\Sigma_{S0,j}$ and $\Sigma_{S1,j}$, respectively.

$$\varphi_{o,j} = \sum_{n=1}^{N} \psi_{n,j} w_n \qquad (6)$$

$$\varphi_{1,j} = \sum_{n=1}^{N} \psi_{n,j} \mu_n w_n \qquad (7)$$

As defined in Eqs. (6) and (7), $\varphi_{o,j}$ and $\varphi_{1,j}$ denote the scalar flux and current, respectively, with $w_n$ being the Gauss-Legendre quadrature weight and $Q_{n,j}$ the internal source for angular direction $n$ in cell $j$. The diamond difference approximation [1] provides the relation between cell-edge and cell-averaged angular fluxes:

$$\psi_{n,j} = \frac{1}{2}\left(\psi_{n,j+\frac{1}{2}} + \psi_{n,j-\frac{1}{2}}\right) \tag{8}$$

Given the source $Q_{n,j}$, Eq. (5) was solved to compute the cell-averaged angular flux $\psi_{n,j}$. The sources $Q_{n,j}$ were generated using Gaussian Random Fields (GRFs) and then used in the fixed source NTE solver to obtain the corresponding angular flux solutions $\psi_{n,j}$. A detailed account of the GRF and its generation methodology is provided in Ref. [17], and readers are referred there for further mathematical and implementation details. The DeepONet and FNO models were trained using paired datasets of anisotropic source inputs $Q(x,\mu)$ and their corresponding angular flux outputs $\psi(x,\mu)$. To enhance scalability, the source term is normalized so that its integral over space and angle represents a unit neutron source. The total source emission density $S(x)$ at position $x$ is obtained by integrating $Q(x,\mu)$ over all angles as in Eq. (9):

$$S(x) = \int_{-1}^{1} Q(x,\mu)d\mu, \text{ with } S_{total} = \int_{0}^{X} S(x)dx \tag{9}$$

In discrete form, these expressions become:

$$S_j = \sum_{n=1}^{N} Q_{n,j} w_n, \ 1 \leq n \leq N \text{ and } S_{total} = \sum_{j=1}^{J} S_j h_j, \ 1 \leq j \leq J \tag{10}$$

A total of 1000 GRF samples of $S(x)$ with mean (M) 0.07, variance ($\sigma^2$) 0.0003, and length scale ($l_S$) 1.2 were generated over 100 spatial points ($h_j = 0.1$ cm) and subsequently normalized according to

$$S_{j,\,nor} = \frac{S_j}{\sum_{j=1}^{100} S_j h_j}, \text{ such that } \sum_{j=1}^{100} S_{j,nor} h_j = 1 \quad (11)$$

For each, $S_{j,\,nor}$, an angle-dependent GRF $Q'(x,\mu)$ with parameters M = 0.5, $\sigma^2 = 0.02$, and $l_Q = 6.0$ was generated with 16 angular bins. These angular sources were then normalized via

$$Q_{n,\,j} = Q'_{n,\,j} \left( \frac{S_{j,\,nor}}{\sum_{n=1}^{16} Q'_{n,j} w_n} \right) \quad (12)$$

ensuring that the consistency condition

$$S_{j,nor} = \sum_{n=1}^{16} Q_{n,\,j} w_n \quad (13)$$

is satisfied with all spatial points $j$. The resulting $Q_{n,j}$ served as model inputs, with $\psi_{n,j}$ from numerical NTE solutions as targets. These input-output pairs were used to train the DeepONet and FNO models, both implemented in PyTorch [24] on a workstation equipped with an NVIDIA RTX 4090 GPU and an Intel i9-14900KS CPU.

Table I summarizes the problem configuration, neural network architectures, and the converged training loss for the surrogate models considered in this work. Three separate models were trained for each surrogate type (FNO and DeepONet), corresponding to scattering ratios $c$ (= $\Sigma_{s0}/\Sigma_t$) of 0.1, 0.5, and 1.0, to highlight distinct physical regimes ranging from absorption dominated to purely scattering transport. The case $c = 0.1$ represents a strongly absorbing medium, $c = 0.5$ corresponds to an intermediate balance between absorption and scattering, and $c = 1.0$ represents a purely scattering regime. For each scattering ratio, both FNO and DeepONet models were trained to a similar level of saturated training loss, using mean squared error (MSE) as the error metric, to ensure a fair comparison.

Table I: Problem Configuration and Training Summary for all models.

| Model ID | Problem Configuration (L = 10 cm) | FNO Architecture | DeepONet Architecture | Training Loss (MSE) |
|---|---|---|---|---|
| $M^{FNO}_{c=0.1}$ & $M^{DeepONet}_{c=0.1}$ | $\Sigma_t = 1.0$ cm$^{-1}$<br>$\Sigma_a = 0.9$ cm$^{-1}$<br>$\Sigma_{s0} = 0.1$ cm$^{-1}$<br>$\Sigma_{s1} = 0.03$ cm$^{-1}$<br>$c = 0.1$ | Latent Space Dimension (Width): 64<br>Number of Fourier Layers: 2<br>Number of Retained Spatial Fourier Modes: 24<br>Number of Retained Angular Fourier Modes: 8<br>Spectral convolution: 2D Fourier convolution<br>Activation Function: ReLU<br>Total Number of Parameters: 3,158,529 | BranchNet: [1600, 1200, 800, 600, 400, 200]<br>TrunkNet: [2, 200, 200, 200, 200, 200]<br>Activation Function = ReLu<br>Total parameters = 3,844,600 | $M^{FNO}_{c=0.1} \sim 10^{-7}$<br>$M^{DeepONet}_{c=0.1} \sim 10^{-7}$ |
| $M^{FNO}_{c=0.5}$ & $M^{DeepONet}_{c=0.5}$ | $\Sigma_t = 1.0$ cm$^{-1}$<br>$\Sigma_a = 0.5$ cm$^{-1}$<br>$\Sigma_{s0} = 0.5$ cm$^{-1}$<br>$\Sigma_{s1} = 0.25$ cm$^{-1}$<br>$c = 0.5$ | | | $M^{FNO}_{c=0.5} \sim 10^{-7}$<br>$M^{DeepONet}_{c=0.5} \sim 10^{-7}$ |
| $M^{FNO}_{c=1.0}$ & $M^{DeepONet}_{c=1.0}$ | $\Sigma_t = 1.0$ cm$^{-1}$<br>$\Sigma_a = 0.0$ cm$^{-1}$<br>$\Sigma_{s0} = 1.0$ cm$^{-1}$<br>$\Sigma_{s1} = 0.25$ cm$^{-1}$<br>$c = 1.0$ | | | $M^{FNO}_{c=1.0} \sim 10^{-7}$<br>$M^{DeepONet}_{c=1.0} \sim 10^{-7}$ |

## II.E. Application to Eigenvalue NTE

The eigenvalue problem is considered in the same 1D slab geometry as the fixed source case, with zero incoming neutron current at both boundaries shown in Figure 4. The slab, 10 cm in length, has spatially uniform absorption, fission, scattering, and total cross sections and contains no external neutron sources; all neutrons arise from fission.

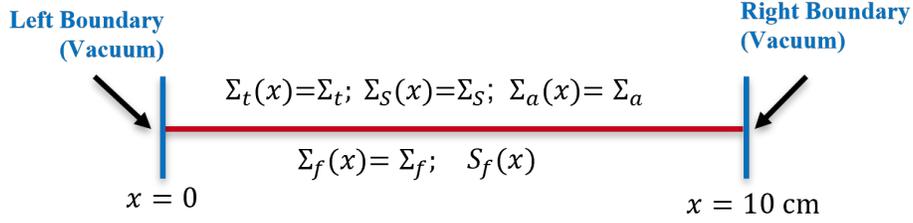

Figure 4. 1D Slab for the k-eigenvalue problem considered in the study.

The steady-state one group NTE governing the eigenvalue problem in the 1D slab is expressed as follows:

$$\mu \frac{\partial \psi(x,\mu)}{\partial x} + \Sigma_t\, \psi(x,\mu) = \int_{-1}^{1} \frac{1}{2}[\Sigma_{S0} + 3\mu\mu' \Sigma_{S1}]\, \psi(x,\mu')d\mu' + \frac{1}{k}\frac{1}{2}\int_{-1}^{1} \vartheta\Sigma_f\, \psi(x,\mu')d\mu';$$

$$0 < x < X,\ -1 \le \mu \le 1 \tag{14}$$

$$\psi(0,\mu) = 0,\ 0 \le \mu \le 1$$

$$\psi(X,\mu) = 0,\ -1 \le \mu \le 0$$

Here, $\Sigma_f$ is the fission cross section, $\vartheta$ the average neutron yield per fission, and $k$ the eigenvalue of Eq. (14). The left-hand side characterizes neutron losses, while the right-hand side comprises the source terms: in-scattering (first term) and fission (second term). To solve numerically, Eq. (14) is discretized in space and angle using the finite difference and $S_N$ methods, respectively, leading to the following form:

$$\frac{\mu_n}{h_j}\left(\psi_{n,j+\frac{1}{2}} - \psi_{n,j-\frac{1}{2}}\right) + \Sigma_{tj}\psi_{n,j} = \frac{1}{2}\left(\Sigma_{S0,j}\varphi_{o,j} + 3\mu_n \Sigma_{S1,j}\varphi_{1,j}\right) + \frac{1}{2}\frac{\vartheta\Sigma_{fj}}{k}\varphi_{o,j}$$

$$1 \le n \le N,\ 1 \le j \le J \tag{15}$$

$$\psi_{n,\frac{1}{2}} = 0,\ \frac{N}{2}+1 \le n \le N, \mu_n > 0$$

$$\psi_{n,J+\frac{1}{2}} = 0, \ 1 \le n \le \frac{N}{2}, \mu_n < 0$$

The symbols in Eq. (15) retain the same meaning as in the fixed source case, with $\Sigma_{fj}$ representing the fission cross section in the $j^{th}$ cell. To ensure that the flux is physically meaningful and uniquely defined, a normalization condition must be imposed. In this study, the following normalization condition is applied:

$$\sum_{j=1}^{J} \vartheta \Sigma_{fj} \varphi_{o,j}^{(l)} h_j = 1 \qquad (16)$$

To numerically solve Eq. (15), we begin with an initial guess

$$k^{(0)} = 1 \text{ and } \varphi_{o,j}^{(0)} = 1 \Big/ \sum_{j=1}^{J} \vartheta \Sigma_{fj} h_j \qquad (17)$$

Then $k^{(l+1)}$ and $\varphi_{o,j}^{(l+1)}$ is recursively defined as follows. Given $k^{(l)}$ and $\varphi_{o,j}^{(l)}$, we compute $k^{(l+1)}$ and $\varphi_{o,j}^{(l+1)}$ by solving a fixed source problem.

$$\frac{\mu_n}{h_j}\left(\psi_{n,j+\frac{1}{2}}^{(l+\frac{1}{2})} - \psi_{n,j-\frac{1}{2}}^{(l+\frac{1}{2})}\right) + \Sigma_{tj}\psi_{n,j}^{(l+\frac{1}{2})} - \frac{1}{2}\left(\Sigma_{S0,j}\varphi_{o,j}^{(l+\frac{1}{2})} + 3\mu_n\Sigma_{S1,j}\varphi_{1,j}^{(l+\frac{1}{2})}\right) = \frac{1}{2}\frac{\vartheta\Sigma_{fj}}{k^{(l)}}\varphi_{o,j}^{(l)}$$

$$1 \le n \le N, \ 1 \le j \le J$$

$$\psi_{n,\frac{1}{2}}^{(l+\frac{1}{2})} = 0, \ \frac{N}{2}+1 \le n \le N, \mu_n > 0 \qquad (18)$$

$$\psi_{n,J+\frac{1}{2}}^{(l+\frac{1}{2})} = 0, \ 1 \le n \le \frac{N}{2}, \mu_n < 0$$

$$\psi_{n,j}^{(l+\frac{1}{2})} = \frac{1}{2}\left(\psi_{n,j+\frac{1}{2}}^{(l+\frac{1}{2})} + \psi_{n,j-\frac{1}{2}}^{(l+\frac{1}{2})}\right), 1 \le n \le N, \ 1 \le j \le J \qquad (19)$$

The cell-averaged angular flux $\psi_{n,j}^{(l+\frac{1}{2})}$ is determined from Eq. (18) via the source iteration method, after which the scalar flux $\varphi_{o,j}^{(l+\frac{1}{2})}$ is computed as

$$\varphi_{o,j}^{(l+\frac{1}{2})} = \sum_{n=1}^{N} \psi_{n,j}^{(l+\frac{1}{2})} w_n, \quad 1 \leq j \leq J \tag{20}$$

The cell average current $\varphi_{1,j}^{(l+\frac{1}{2})}$ is calculated similarly using Eq. (21)

$$\varphi_{1,j}^{(l+\frac{1}{2})} = \sum_{n=1}^{N} \mu_n \psi_{n,j}^{(l+\frac{1}{2})} w_n, \quad 1 \leq j \leq J \tag{21}$$

The $k^{(l+1)}$ is then calculated as follows:

$$k^{(l+1)} = k^{(l)} \sum_{j=1}^{J} \vartheta \Sigma_{fj} \varphi_{o,j}^{(l+\frac{1}{2})} h_j \tag{22}$$

Using the $\varphi_{o,j}^{(l+\frac{1}{2})}$ from Eq. (20) and $k^{(l+1)}$ from Eq. (22), $\varphi_{o,j}^{(l)}$ is calculated from Eq. (23).

$$\varphi_{o,j}^{(l)} = \frac{k^{(l)}}{k^{(l+1)}} \varphi_{o,j}^{(l+\frac{1}{2})} \tag{23}$$

This process continues until the convergence criterion is met.

In this standard $S_N$ approach for the k-eigenvalue neutron transport problem, the inner iteration loop, which updates the scalar flux via repeated transport sweeps, dominates computational cost. To significantly accelerate this process, we substitute this loop with DeepONet or FNO models trained on fixed source problems. During each power iteration, the models predict the angular flux directly from the current source, bypassing iterative transport solves. The resulting flux is used to update both the eigenvalue and fission source, with iterations continuing until convergence.

Consequently, the k-eigenvalue solution becomes much more efficient, since the computationally intensive transport solve in each power iteration is no longer required.

## III. RESULTS AND DISCUSSION

### III.A. Fixed Source Problem

*III.A.1. Performance Evaluation Using Anisotropic Source*

To assess the predictive capability and computational efficiency of the DeepONet and FNO models, six test cases were generated following the procedure described in Section II.D by varying $l_S$ of $S(x)$ and $l_Q$ of $Q'(x,\mu)$ as shown in Table II. Case 1 corresponds to the same correlation lengths $l_S$ and $l_Q$ used during training. The remaining cases introduce variations in either $l_S$, $l_Q$, or both, enabling an assessment of model robustness to changes in input function correlation structure.

Table II. Test Cases Using Anisotropic Source.

| Case ID | Description |
|---------|-------------|
| Case 1 | [ $S(x): l_S = 1.2;\ Q'(x,\mu): l_Q = 6.0$ ] |
| Case 2 | [ $S(x): l_S = 1.2;\ Q'(x,\mu): l_S = 4.0$ ] |
| Case 3 | [ $S(x): l_S = 1.2;\ Q'(x,\mu): l_S = 8.0$ ] |
| Case 4 | [ $S(x): l_S = 1.0;\ Q'(x,\mu): l_S = 6.0$ ] |
| Case 5 | [ $S(x): l_S = 1.4;\ Q'(x,\mu): l_S = 6.0$ ] |
| Case 6 | [ $S(x): l_S = 1.4;\ Q'(x,\mu): l_S = 4.0$ ] |

For each configuration, one hundred test samples were evaluated. The average MSE between the predicted and reference angular fluxes over 100 samples served as the primary accuracy metric. Furthermore, since the scalar flux is a physically observable quantity and is commonly employed in practical applications, the average of the mean relative error (MRE) of the scalar flux is reported to assess model performance. These error metrics are calculated as per definitions given in the Appendix. To quantify the computational speedup, the mean inference time of each model was recorded and expressed as a percentage of the runtime of the traditional $S_N$ solver. These performance metrics for the models are reported in Tables III-V.

Table III. Performance of the Models with $c = 0.1$.

| Cases | Average MSE of Angular Flux | | Average MRE of Scalar Flux (%) | | Runtime (% of $S_N$ Solver) | |
|---|---|---|---|---|---|---|
| | $M^{FNO}_{c=0.1}$ | $M^{DeepONet}_{c=0.1}$ | $M^{FNO}_{c=0.1}$ | $M^{DeepONet}_{c=0.1}$ | $M^{FNO}_{c=0.1}$ | $M^{DeepONet}_{c=0.1}$ |
| Case 1 | $6.43 \times 10^{-8}$ | $1.39 \times 10^{-6}$ | 0.118 | 1.253 | 11.27 | 5.87 |
| Case 2 | $2.27 \times 10^{-7}$ | $2.14 \times 10^{-6}$ | 0.193 | 1.393 | 11.95 | 6.15 |
| Case 3 | $4.46 \times 10^{-8}$ | $1.23 \times 10^{-6}$ | 0.092 | 1.207 | 12.21 | 6.26 |
| Case 4 | $5.93 \times 10^{-8}$ | $2.11 \times 10^{-6}$ | 0.117 | 1.605 | 12.06 | 6.17 |
| Case 5 | $6.49 \times 10^{-8}$ | $1.02 \times 10^{-6}$ | 0.118 | 1.052 | 12.11 | 6.29 |
| Case 6 | $2.26 \times 10^{-7}$ | $1.76 \times 10^{-6}$ | 0.193 | 1.209 | 11.98 | 6.14 |

Table III summarizes the performance of the surrogate models for the strongly absorbing case $c = 0.1$. Overall, the FNO model achieves an average MSE on the order of $10^{-7}$-$10^{-8}$ while the DeepONet model exhibits MSE values on the order of $10^{-6}$. In terms of scalar flux accuracy, the FNO model consistently maintains an average MRE below approximately 0.2%, whereas the DeepONet model yields a maximum MRE value of 1.605%. From a computational perspective, both models provide significant speedups relative to the reference $S_N$ solver, with FNO runtimes remaining below about 13% of the $S_N$ cost and DeepONet achieving even lower runtimes, typically below 7%. These results indicate that, for $c = 0.1$, FNO offers higher predictive accuracy, while DeepONet provides greater computational efficiency.

Table IV. Performance of the Models with $c = 0.5$.

| Cases | Average MSE of Angular Flux | | Average MRE of Scalar Flux (%) | | Runtime (% of $S_N$ Solver) | |
|---|---|---|---|---|---|---|
| | $M^{FNO}_{c=0.5}$ | $M^{DeepONet}_{c=0.5}$ | $M^{FNO}_{c=0.5}$ | $M^{DeepONet}_{c=0.5}$ | $M^{FNO}_{c=0.5}$ | $M^{DeepONet}_{c=0.5}$ |
| Case 1 | $9.68 \times 10^{-8}$ | $3.63 \times 10^{-6}$ | 0.087 | 1.006 | 4.69 | 2.41 |
| Case 2 | $3.33 \times 10^{-7}$ | $4.62 \times 10^{-6}$ | 0.140 | 1.118 | 4.87 | 2.48 |
| Case 3 | $6.75 \times 10^{-8}$ | $3.42 \times 10^{-6}$ | 0.068 | 0.965 | 4.76 | 2.44 |
| Case 4 | $8.12 \times 10^{-8}$ | $4.32 \times 10^{-6}$ | 0.086 | 1.179 | 4.90 | 2.51 |
| Case 5 | $9.67 \times 10^{-8}$ | $2.91 \times 10^{-6}$ | 0.087 | 0.883 | 4.66 | 2.40 |
| Case 6 | $3.25 \times 10^{-7}$ | $3.90 \times 10^{-6}$ | 0.140 | 1.005 | 4.78 | 2.48 |

Table IV presents the performance metrics of the surrogate models for the intermediate scattering scenario $c = 0.5$. The FNO model achieves angular flux MSE values on the order of $10^{-7}$-$10^{-8}$, whereas the DeepONet model exhibits slightly higher MSE, around $10^{-6}$. Regarding scalar flux predictions, FNO maintains a very low average MRE, below roughly 0.15%, while DeepONet's MRE is nearly 1%. In terms of computational cost, both surrogates offer notable reductions relative to the $S_N$ solver, with FNO requiring less than 5% of the original runtime and while for DeepONet it is <2.5%. As observed for the $c = 0.1$ models, these findings indicate that, for $c = 0.5$, FNO achieves relatively greater accuracy, while DeepONet remains advantageous in terms of computational efficiency.

Table V. Performance of the Models with $c = 1.0$.

| | Average MSE | Average MRE | Runtime |
|---|---|---|---|

| Cases | of Angular Flux | | of Scalar Flux (%) | | (% of $S_N$ Solver) | |
|---|---|---|---|---|---|---|
| | $M_{c=1.0}^{FNO}$ | $M_{c=1.0}^{DeepONet}$ | $M_{c=1.0}^{FNO}$ | $M_{c=1.0}^{DeepONet}$ | $M_{c=1.0}^{FNO}$ | $M_{c=1.0}^{DeepONet}$ |
| Case 1 | $3.22 \times 10^{-6}$ | $6.18 \times 10^{-5}$ | 0.032 | 0.267 | 0.211 | 0.110 |
| Case 2 | $5.32 \times 10^{-6}$ | $6.79 \times 10^{-5}$ | 0.046 | 0.282 | 0.209 | 0.110 |
| Case 3 | $2.76 \times 10^{-6}$ | $6.03 \times 10^{-5}$ | 0.026 | 0.262 | 0.207 | 0.111 |
| Case 4 | $3.26 \times 10^{-6}$ | $5.39 \times 10^{-5}$ | 0.031 | 0.280 | 0.212 | 0.109 |
| Case 5 | $3.96 \times 10^{-6}$ | $7.13 \times 10^{-5}$ | 0.032 | 0.272 | 0.210 | 0.110 |
| Case 6 | $6.00 \times 10^{-6}$ | $8.04 \times 10^{-5}$ | 0.047 | 0.289 | 0.210 | 0.110 |

Table V reports the performance of the surrogate models for the purely scattering case $c = 1.0$. Both FNO and DeepONet show slightly higher MSE than in the $c = 0.1$ and $c = 0.5$ cases. Scalar flux predictions remain highly accurate for both models. In terms of runtime, both surrogates are extremely efficient, with DeepONet being the fastest. Overall, FNO continues to offer marginally better accuracy, while DeepONet excels in computational efficiency, demonstrating the complementary strengths of the two approaches in a scattering dominated regime.

After evaluating the models on the six discrete anisotropic source cases listed in Table II, we further tested their generalization capability across the continuous $(l_S, l_Q)$ parameter plane. A total of 500 random points were sampled within the $(l_S, l_Q)$ domain, and corresponding source functions were generated for each sample. Both DeepONet and FNO were then used to predict the angular flux for these sources, and the MSE between the predicted and reference fluxes was computed for each $(l_S, l_Q)$ point. The resulting MSE distributions for DeepONet and FNO are visualized in Figures 5-7, providing a comprehensive overview of model accuracy across a wide range of source correlation lengths. As expected, the MSE remains low for points close to the original training configurations. However, as the testing points move farther away from the training locations in the

$(l_S, l_Q)$ plane, the MSE gradually increases. Despite this, both models maintain a lower MSE across a large parameter space.

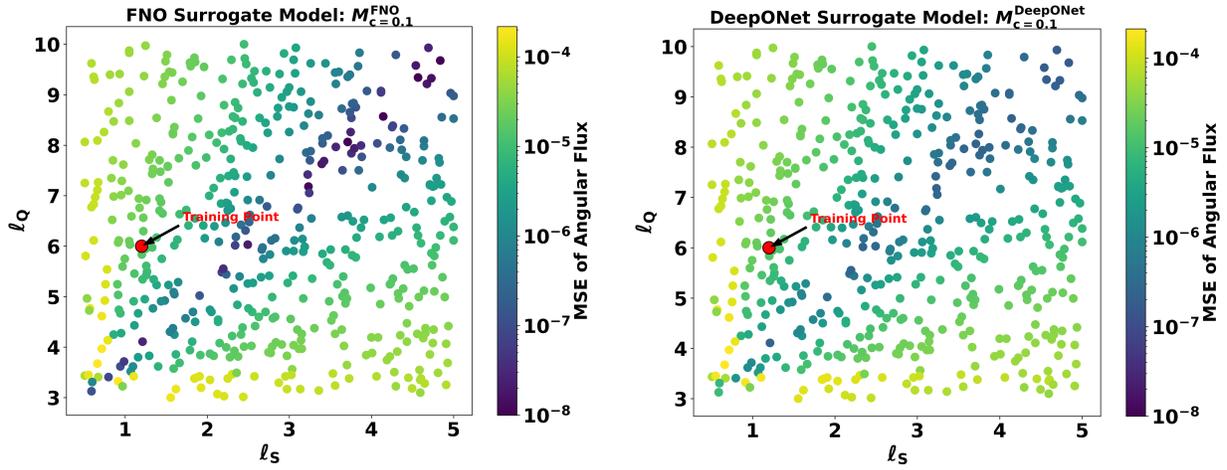

Figure 5. MSE of angular flux for models with $c = 0.1$ on the $(l_S, l_Q)$ plane.

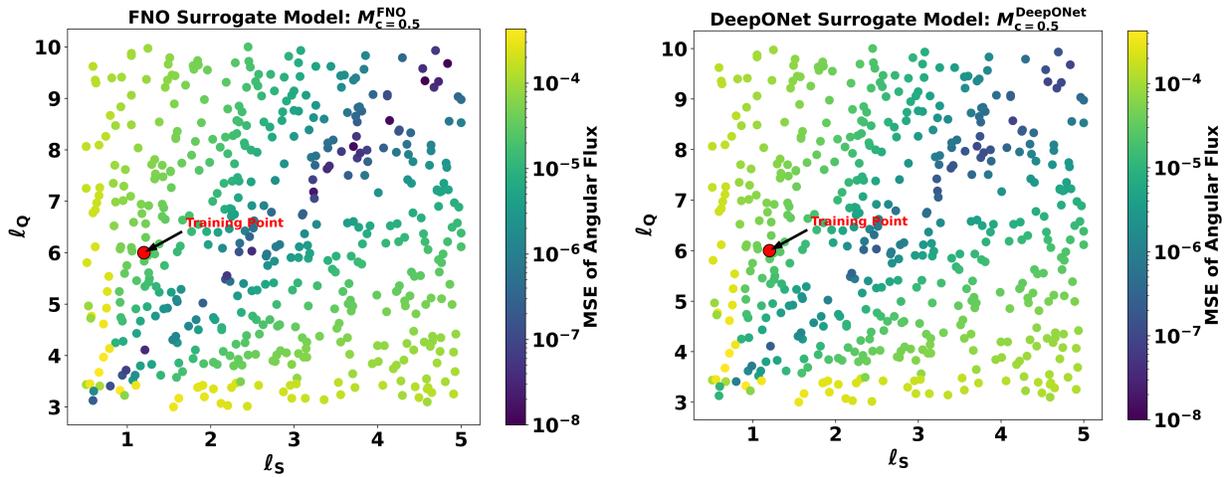

Figure 6. MSE of angular flux for models with $c = 0.5$ on the $(l_S, l_Q)$ plane.

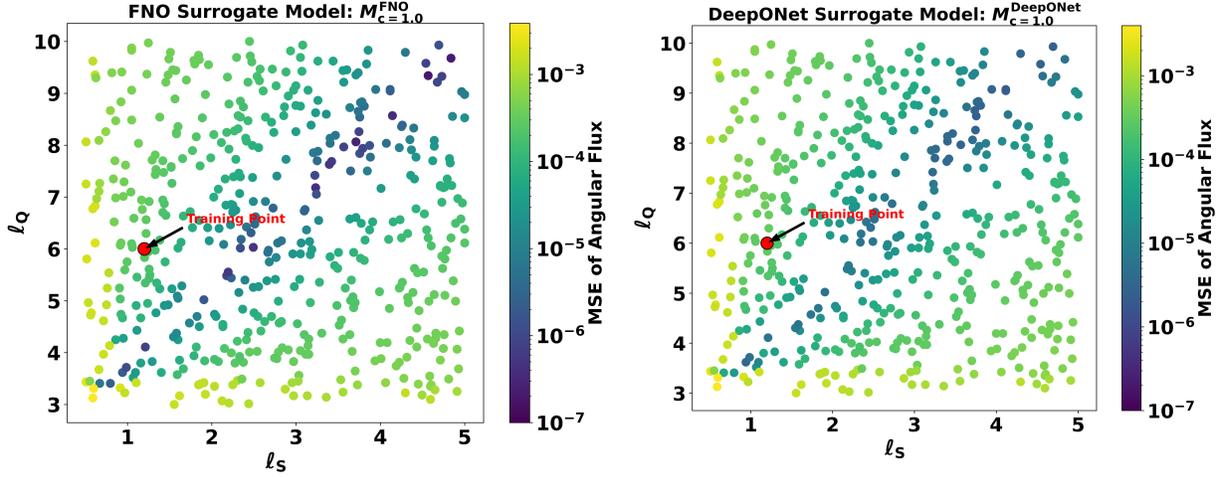

Figure 7. MSE of angular flux for models with $c = 1.0$ on the $(l_S, l_Q)$ plane.

*III.A.2. Performance Evaluation Using Isotropic Source*

To evaluate model behavior under simplified angular dependence, both DeepONet and FNO were tested using an isotropic source. The normalized spatial source $S_{j,\,nor}$ was first computed using Eq. (11) and subsequently used to form the source term $Q_{n,\,j} = \frac{S_{j,\,nor}}{2}$. Two test cases were generated by varying the spatial correlation length $l_S$, and the corresponding performance metrics are summarized in Table VI. Overall, both FNO and DeepONet exhibit a similar type of behavior as observed for the anisotropic cases, with FNO achieving relatively higher accuracy for both angular and scalar flux predictions. This trend is consistent across all models with different $c$ values. The results indicate that the models maintain their good performance regardless of source isotropy.

Table VI. Model Performance for the Isotropic Source.

| Cases | Average MSE of Angular Flux | | Average MRE of Scalar Flux (%) | |
|---|---|---|---|---|
| | $M^{FNO}_{c=0.1}$ | $M^{DeepONet}_{c=0.1}$ | $M^{FNO}_{c=0.1}$ | $M^{DeepONet}_{c=0.1}$ |
| Case 1 ($l_S = 1.6$) | $2.51 \times 10^{-8}$ | $5.06 \times 10^{-7}$ | 0.028 | 0.804 |
| Case 2 ($l_S = 0.8$) | $1.85 \times 10^{-8}$ | $3.62 \times 10^{-6}$ | 0.057 | 2.217 |
| Cases | $M^{FNO}_{c=0.5}$ | $M^{DeepONet}_{c=0.5}$ | $M^{FNO}_{c=0.5}$ | $M^{DeepONet}_{c=0.5}$ |
| Case 1 ($l_S = 1.6$) | $4.06 \times 10^{-8}$ | $1.86 \times 10^{-6}$ | 0.027 | 0.656 |
| Case 2 ($l_S = 0.8$) | $4.10 \times 10^{-8}$ | $8.00 \times 10^{-6}$ | 0.059 | 1.702 |
| Cases | $M^{FNO}_{c=1.0}$ | $M^{DeepONet}_{c=1.0}$ | $M^{FNO}_{c=1.0}$ | $M^{DeepONet}_{c=1.0}$ |
| Case 1 ($l_S = 1.6$) | $3.45 \times 10^{-6}$ | $5.84 \times 10^{-5}$ | 0.022 | 0.251 |
| Case 2 ($l_S = 0.8$) | $1.46 \times 10^{-6}$ | $5.58 \times 10^{-5}$ | 0.020 | 0.291 |

### III. B. Eigenvalue Problem

For the eigenvalue problem, the same one dimensional slab geometry was used as in the fixed source case, with $\Sigma_t = 1.0$ cm$^{-1}$, $\Sigma_{s0} = 0.5$ cm$^{-1}$, and $\Sigma_{s1} = 0.25$ cm$^{-1}$. Based on these cross section values, the fixed source model with c = 0.5 ($M^{FNO}_{c=0.5}$ or $M^{DeepONet}_{c=0.5}$) was integrated into the $S_N$ eigenvalue solver to replace the inner transport sweep loop. Several cases were then analyzed, each corresponding to a different fission cross section (Case 1: $\Sigma_f = 0.2$ cm$^{-1}$, Case 2: $\Sigma_f = 0.25$ cm$^{-1}$ and Case 3: $\Sigma_f = 0.3$ cm$^{-1}$ with $\vartheta = 2.6$ for each case). In addition to different $\Sigma_f$, each of the cases were evaluated for different space and angular grid. The results obtained using the DeepONet and FNO based solvers were compared with those from the traditional $S_N$ transport solver. The same convergence criteria of $10^{-5}$ were applied to both the eigenvalue and scalar flux within the power iteration loop for all three approaches.

Table VII: Eigenvalue for Different Test Cases.

| Cases | Mesh Size | $k_{eff}$ | | | $\Delta k_{eff}$ (pcm) | |
|---|---|---|---|---|---|---|
| | | Traditional $S_N$ Solver | DeepONet Based Solver | FNO Based Solver | DeepONet Based Solver | FNO Based Solver |
| Case 1 | $J=100, N=16$ | 0.98001 | 0.97909 | 0.97967 | -92.0 | -34.0 |
| Case 1 | $J=1000, N=32$ | 0.98003 | 0.97911 | 0.98075 | -92.0 | 72.0 |
| Case 2 | $J=100, N=16$ | 1.22500 | 1.22385 | 1.22459 | -115.0 | -41.0 |
| Case 2 | $J=1000, N=32$ | 1.22502 | 1.22389 | 1.22593 | -113.0 | 91.0 |
| Case 3 | $J=100, N=16$ | 1.46997 | 1.46862 | 1.46951 | -135.0 | -46.0 |
| Case 3 | $J=1000, N=32$ | 1.47000 | 1.46866 | 1.47112 | -134.0 | 112.0 |

Table VII presents the eigenvalue predictions for three test cases, each corresponding to a $\Sigma_f$, evaluated using both relatively coarse ($J = 100$, $N = 16$) and fine ($J = 1000$, $N = 32$) discretization. The variations in $\Sigma_f$ produce progressively larger eigenvalues, enabling assessment of surrogate model performance across a wider reactivity range. Across all $\Sigma_f$ cases and mesh resolutions, the neural operator based solvers demonstrate good agreement with the traditional $S_N$ reference. For the relatively coarse discretization, both solvers underpredict the eigenvalue, with the DeepONet based solver showing deviations between 92 pcm and 135 pcm, while the FNO based solver exhibits smaller deviations, ranging from 34 pcm to 46 pcm. When the mesh is refined to ($J = 1000$, $N = 32$), DeepONet maintains nearly identical deviations as in the coarse grid, again underpredicting the eigenvalue in all cases. In contrast, the FNO based solver slightly overpredicts the eigenvalue on the refined mesh, with errors ranging from 72 to 112 pcm. Although these deviations are larger than those observed on the coarse mesh, they remain moderate and preserve the correct ordering of eigenvalues across the three fission cross section cases. These findings underscore a significant advantage of operator learning approaches: by capturing the underlying functional relationship between the source and flux, the models can generalize reliably to discretization not encountered during training. This ability is particularly beneficial for accelerating eigenvalue computations, where mesh refinement is often necessary to attain highly accurate results.

Figure 8 shows a comparison of the predicted scalar flux with the $S_N$ solution. To quantify the deviation, MRE of scalar flux was calculated. Table VIII summarizes the MRE of the scalar flux and the corresponding computational cost for the eigenvalue problem across different spatial and angular mesh configurations. For all cases, the DeepONet and FNO based solvers achieve errors of less than 1%, demonstrating strong agreement with the reference $S_N$ solution. Across the relatively coarse discretization ($J = 100$, $N = 16$), the FNO generally attains slightly lower flux errors than DeepONet, whereas the performance gap narrows for the finer meshes ($J = 1000$, $N = 32$).

Table VIII: Relative $L_2$ Error of Scalar Flux and Computational Cost for the Eigenvalue Problem.

| Cases | Mesh Size | MRE of Scalar Flux (%) | | Run Time (% of Traditional $S_N$ Solver) | |
|---|---|---|---|---|---|
| | | DeepONet Based Solver | FNO Based Solver | DeepONet Based Solver | FNO Based Solver |
| Case 1 | $J = 100$, $N = 16$ | 0.737 | 0.116 | 0.761 | 1.103 |
| Case 1 | $J = 1000$, $N = 32$ | 0.738 | 0.581 | 0.062 | 0.056 |
| Case 2 | $J = 100$, $N = 16$ | 0.736 | 0.117 | 0.855 | 1.240 |
| Case 2 | $J = 1000$, $N = 32$ | 0.738 | 0.579 | 0.068 | 0.058 |
| Case 3 | $J = 100$, $N = 16$ | 0.737 | 0.118 | 0.969 | 1.357 |
| Case 3 | $J = 1000$, $N = 32$ | 0.739 | 0.576 | 0.075 | 0.070 |

In terms of computational efficiency, both neural operator solvers achieve dramatic reductions in run time compared to the traditional $S_N$ solver. Notably, on the finer mesh, the DeepONet and FNO solvers execute at <0.1% of the traditional computation time, underscoring their substantial acceleration for high-resolution eigenvalue transport calculations. Overall, the results demonstrate that DeepONet and FNO provide accurate and efficient surrogate models for k-eigenvalue neutron transport problems across varying fission cross sections and grid resolutions. These findings reinforce the potential of neural operator methodologies as powerful acceleration tools for reactor physics simulations.

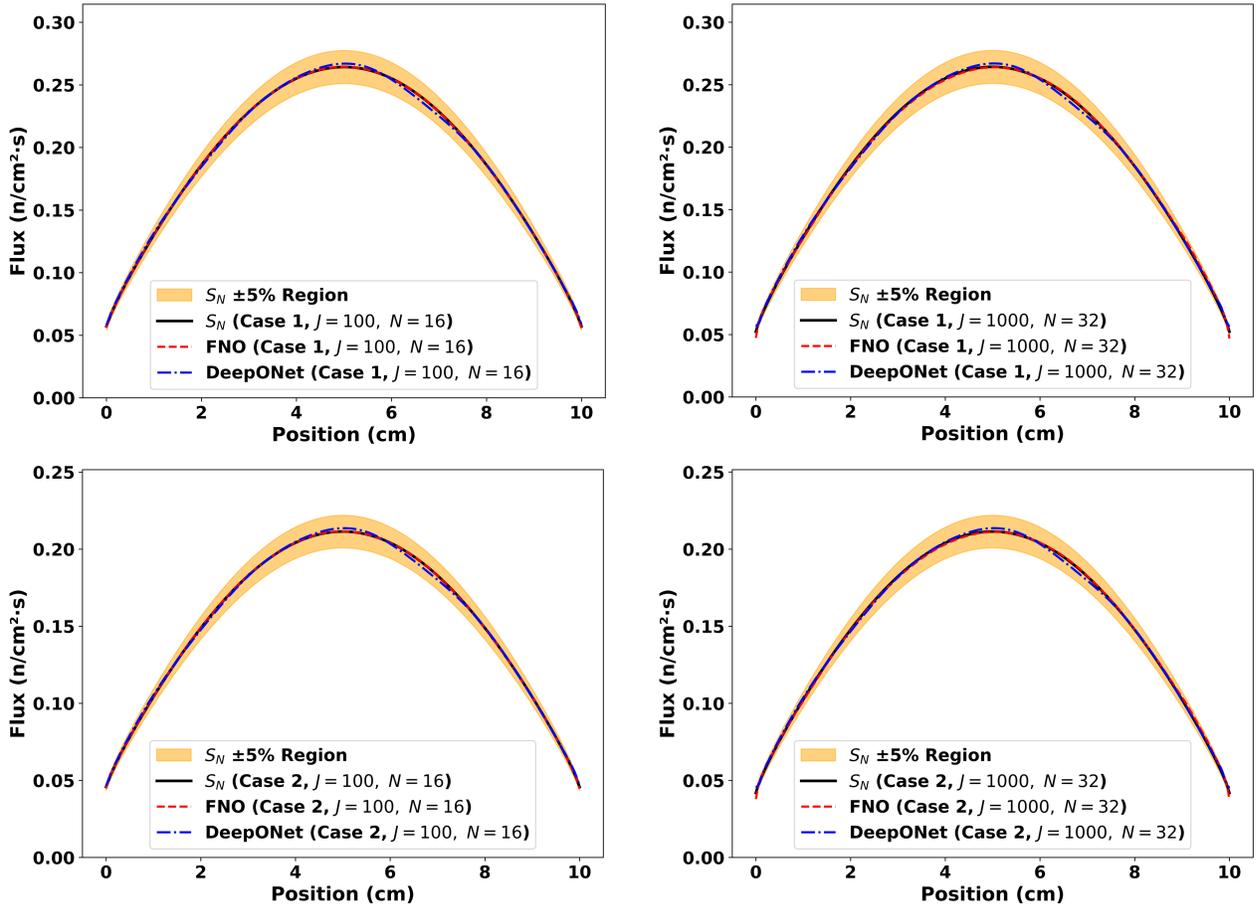

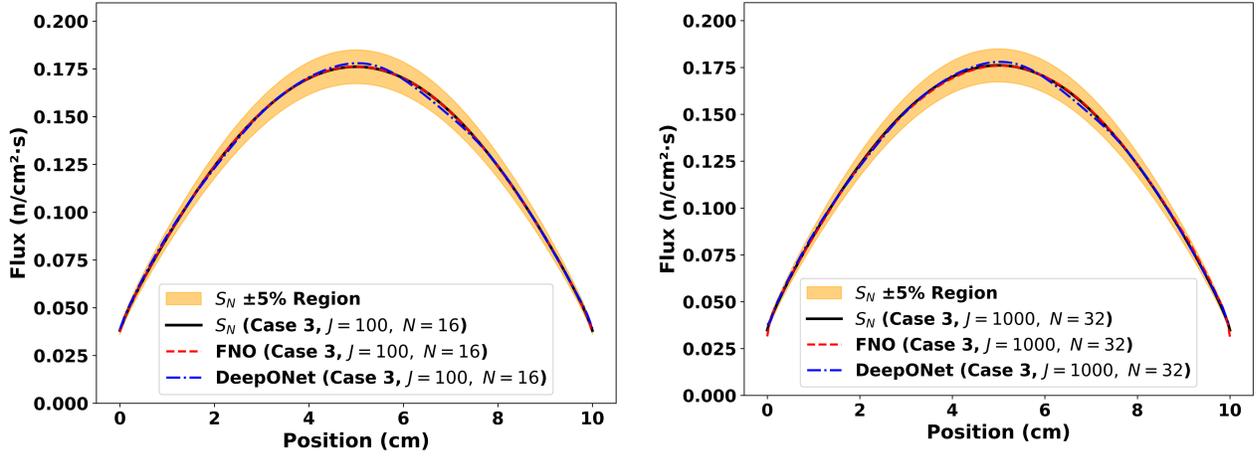

Figure 8. Scalar Flux distribution for eigenvalue problem.

## IV. CONCLUSIONS

Neural operator based surrogate models were investigated for fixed source and k-eigenvalue neutron transport problems in a one-dimensional slab geometry. The results show that both DeepONet and FNO can approximate the mapping from anisotropic neutron sources to angular fluxes with good accuracy while significantly reducing computational cost relative to conventional $S_N$ solvers. In fixed source calculations, FNO generally achieved lower prediction errors, whereas DeepONet exhibited slightly higher errors but improved computational efficiency. In k-eigenvalue calculations, the replacement of transport sweeps loop with neural operator evaluations resulted in substantial reductions in runtime. The FNO based solver showed closer agreement with reference eigenvalues, while the DeepONet based solver exhibited slightly larger deviations. Nevertheless, both approaches produced eigenvalues within reasonable accuracy for practical applications. These results indicate that neural operator surrogates can be a viable and efficient alternative to traditional transport solvers. By directly learning the transport solution operator, the proposed framework enables rapid evaluations that are well suited for repeated analyses, uncertainty studies, and design-oriented calculations. Future work will focus on extending the proposed methodology to higher-dimensional geometries and on developing approaches for solving NTE that enables a single model to generalize across a broad range of cross section values, thereby supporting more realistic reactor physics applications. Moreover, the proposed fast surrogate models have the potential to significantly accelerate existing reduced-order modeling frameworks, such as proper generalized decomposition (PGD), by serving as efficient solvers for the low-dimensional transport equations arising in PGD formulations, as demonstrated in prior PGD studies [25-27].

**APPENDIX**

**A. Error Metrics**

The mean squared error (MSE) of the angular flux was calculated using Eq. (A1)

$$\text{MSE} = \frac{1}{JN} \sum_{j=1}^{J} \sum_{n=1}^{N} (\psi_{nj}^{Tr} - \psi_{nj}^{Pr})^2 \tag{A1}$$

where, $\psi_{nj}^{Tr}$ and $\psi_{nj}^{Pr}$ represents the true value and predicted value of the angular flux, respectively. To evaluate the accuracy of the scalar predictions, the point wise relative error ($\text{RE}_i$) for the $i^{th}$ data point was computed using the following formula

$$\text{RE}_i\ (\%) = \frac{|\varphi_{0,j}^{Tr} - \varphi_{0,j}^{Pr}|}{\varphi_{0,j}^{Tr}} \times 100 \tag{A2}$$

Here, $\varphi_{0,j}^{Tr}$ and $\varphi_{0,j}^{Pr}$ represents the true value and predicted value of the scalar flux for the respectively. The mean relative error (MRE) for all the observation points is then calculated as

$$\text{MRE}\ (\%) = \frac{1}{J} \sum_{i=1}^{J} \text{RE}_i\ (\%) \tag{A3}$$

## B. Representative Plot of the Source and Prediction Error

To illustrate the angular flux prediction error in the $(x, \mu)$ plane, Figure B1 presents the anisotropic source realization yielding the minimum MSE and the corresponding error fields for models with c = 0.5 for Case 1 and Case 6, which are shown as representative in-distribution and out-of-distribution test cases, respectively. Case 1 uses the same $l_S$ and $l_Q$ as the training data, whereas in Case 6 both differ from those used in training (Table II). Figure B2 shows a similar plot for isotropic source for the same models (c = 0.5).

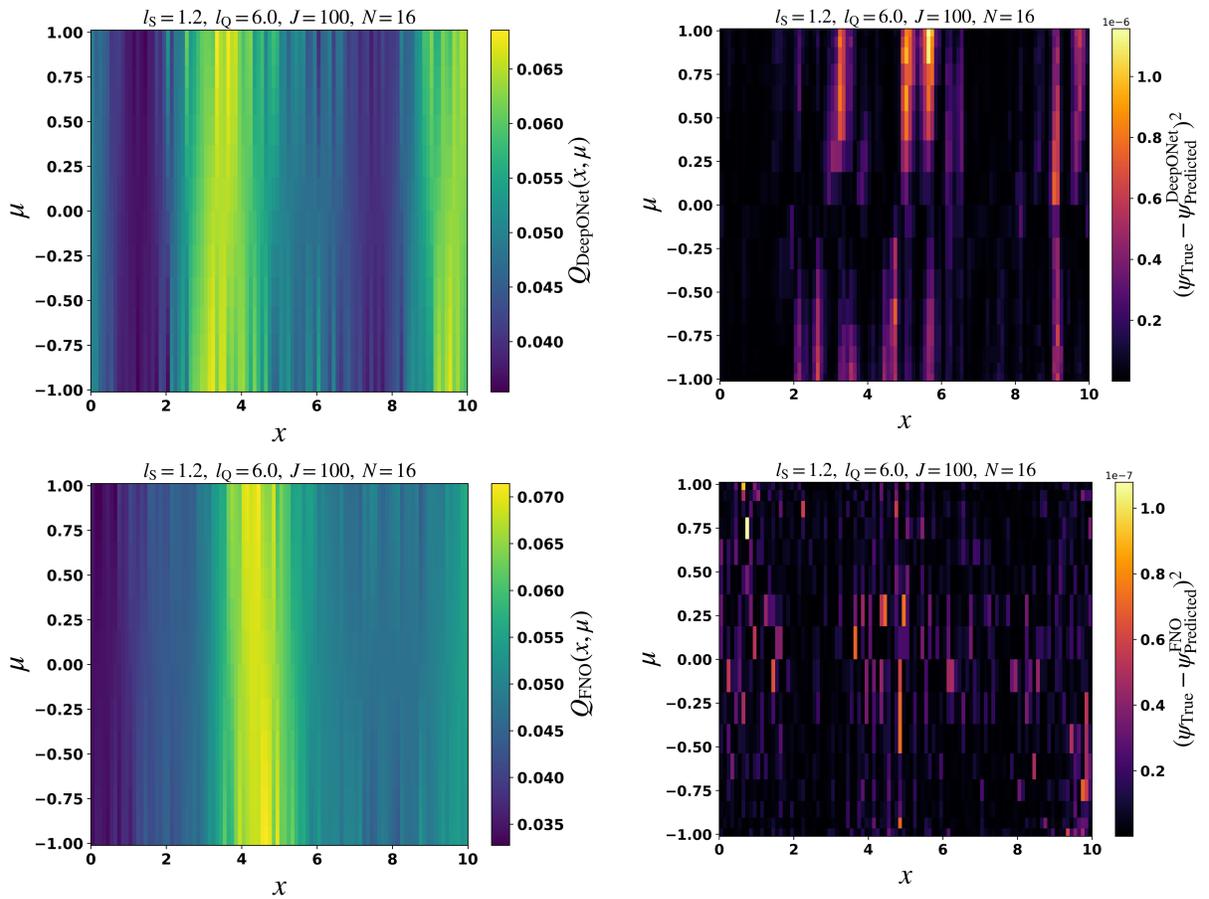

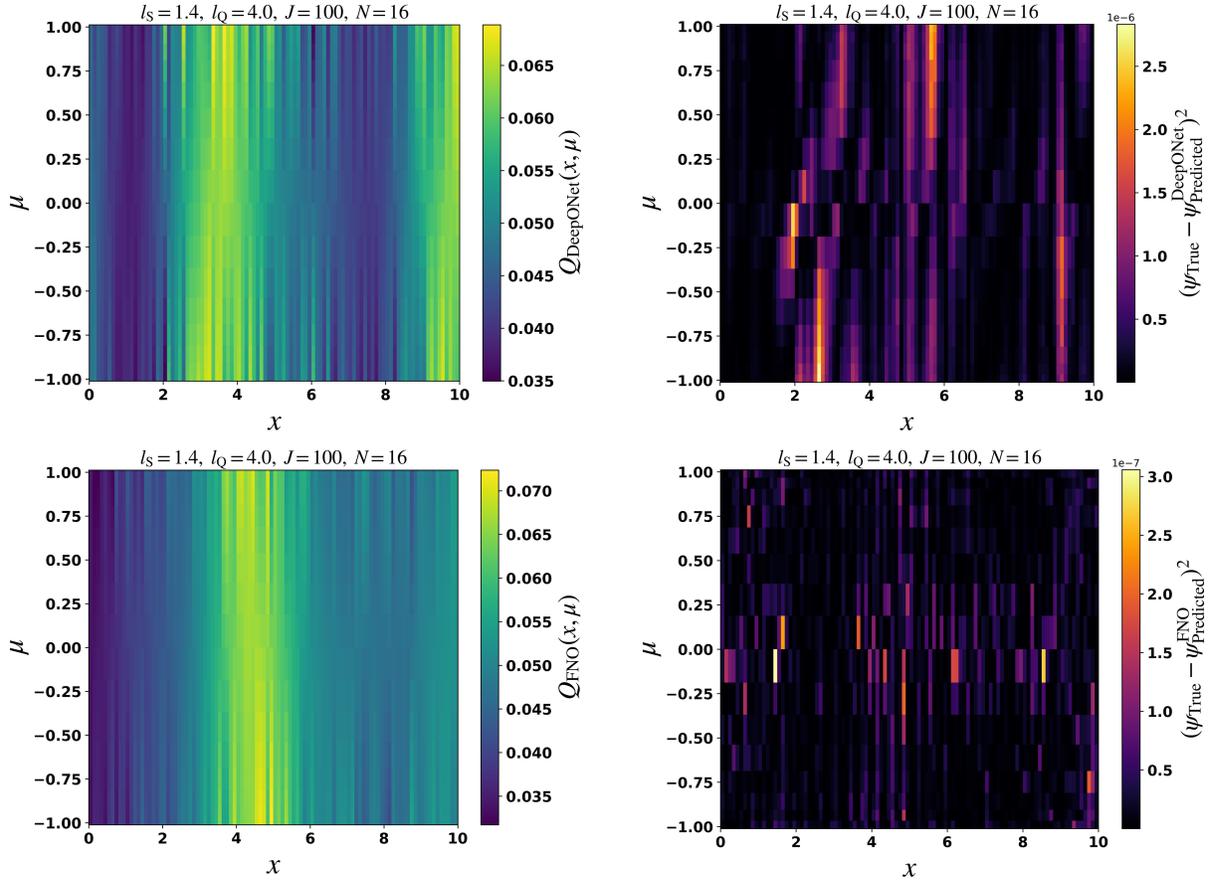

Figure B1. Source (Anisotropic) realizations and corresponding angular flux prediction errors for model with c = 0.5.

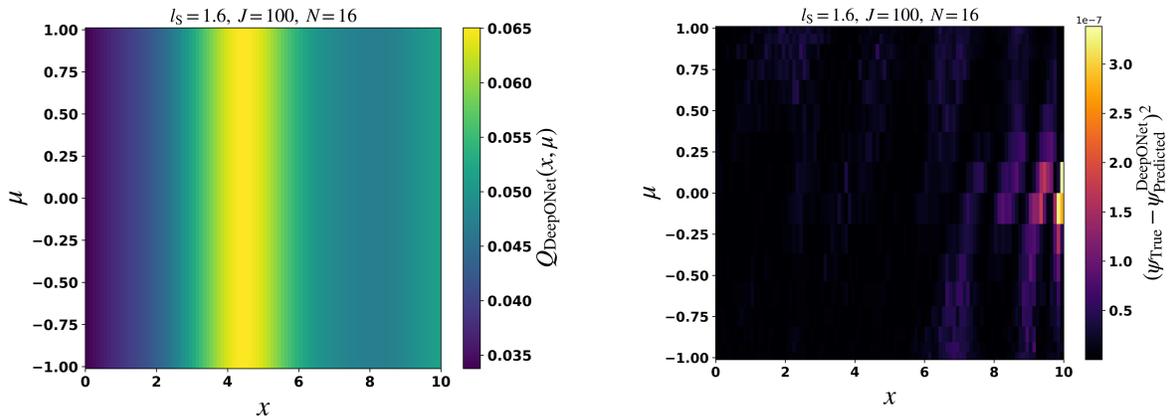

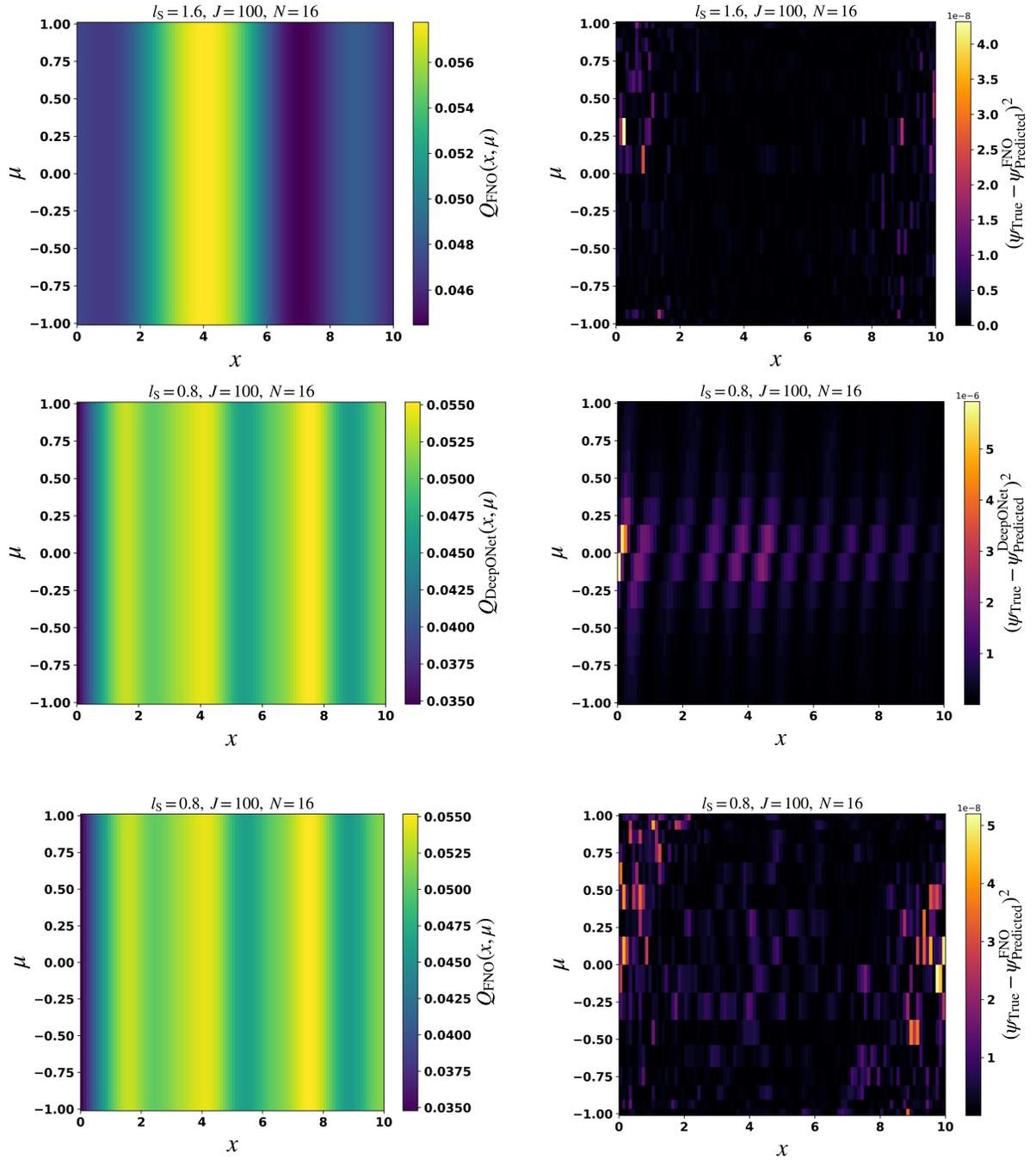

Figure B2. Source (isotropic) realizations and corresponding angular flux prediction errors for model with c = 0.5.

**Disclosure Statement**

The authors report there are no competing interests to declare.